\documentclass[a4paper,10pt]{article}
\usepackage[utf8x]{inputenc}

\title{  Noether's Charge in the Super-Group Field Cosmology }
\author{ Mir Faizal \\
Mathematical Institute, University of Oxford
\\ Oxford
OX1 3LB, United Kingdom  
 }
\date{}
\begin{document}

\maketitle

\begin{abstract} In this paper we  will analyse the third quantization of a model 
of super-group field cosmology with gauge symmetry. 
The effect of creation and annihilation of bosonic and fermionic  
 universes in the multiverse will also be analysed. 
We will also construct a third quantized Noether's charge which 
 will be conserved even when the number of universes is not conserved. 
Finally, we will construct both the third quantized BRST and the third quantized anti-BRST charges 
for this theory
and use them to show that the creation and annihilation of  universes is a unitarity process. 
\end{abstract}
\section{Introduction}
In loop quantum gravity  the Hamiltonian constraint  is expressed in terms of 
Ashtekar-Barbero connection and densitized triad \cite{1}-\cite{1a}. The 
 curvature of the connection is 
expressed through 
the holonomy around a loop. The area operator 
in loop quantum gravity has a discrete spectrum and so it cannot take 
arbitrary small nonzero values. 
Furthermore, in loop quantum gravity has been studied using spin networks \cite{s}-\cite{sa}  and  spin foam models \cite{f}-\cite{fa}. 
Spin foam models are generated by the time evolution of these spin networks. 

Despite the success of loop quantum gravity, the process that change the topology of spacetime cannot be studied in it. 
This is because loop quantum gravity is a second quantized formalism, and just like the first quantized formalism 
cannot be used to analyse the situation when the particle number is not conserved, the second quantized formalism 
cannot be used to analyse the situation where the topology changes dynamically. 
To analyse such processes we have to resort to a third quantized formalism
 \cite{t}-\cite{ta}. 

Third quantization of loop quantum gravity is called 
 group fields theory, and it is a field theory on a group manifold \cite{g}-\cite{ga}.
However, at present this theory is not fully developed and understood. On the other  hand 
minisuperspace approximations to loop quantum gravity called 
  loop quantum cosmology  are better understood \cite{c}-\cite{ca}.
The third quantization of such models is called 
group field cosmology \cite{m}-\cite{ma}. Recently, a supersymmetric generalization of this loop 
quantum cosmology has also been made \cite{sgfc}. In the model thus, developed gauge symmetry was also incorporated.
The BRST symmetry of this gauge invariant super-group field cosmology was also studied. 

It is known that for any theory with a gauge symmetry, there are unphysical degrees of freedom. 
These unphysical degrees of freedom have to be removed before quantizing the theory. 
This is done by fixing a gauge. The gauge fixing is incorporated at the quantum level by 
adding a gauge fixing and a ghost term. The action thus obtained is invariant under 
a new symmetry called the BRST symmetry \cite{brst}-\cite{brst1}. It is also invariant under another symmetry called 
the anti-BRST symmetry \cite{abrst}-\cite{abrst1}. 
 In conventional gauge theories the BRST and the anti-BRST symmetries have been studied in 
various gauge \cite{gauge}-\cite{gauge1}. As the super-group field cosmology has been found to possess a gauge symmetry, 
in this paper we will  analyse the BRST and the anti-BRST symmetries for it in  Landau  gauge. 
We will also construct a third quantized Noether's charge, and show that this charge will be conserved in the 
multiverse, even though the number of universes is not conserved.

\section{Third Quantized Noether's theorem}
In this section we will construct a third quantized Noether's theorem. 
In a multiverse different universes can collide and get annihilated. Similarly, new universes can form from the collision of the 
previous universe. So, just like in a second quantized interacting field theory the particles number is not conserved, 
in a third quantized loop quantum gravity, the number of universes is not conserved. However,  
in a second quantized field theory, conserved charges can be constructed by using the Noether's theorem. 
This charges correspond to  symmetries under which a second quantized field theory action is invariant. 
In a third quantized theory, we can also construct a third quantized version of the Noether's theorem. 
This will also correspond to symmetries under which a third quantized field theory action is invariant. 
Thus, in a multiverse, even after the creation of new universes these quantities will be conserved.
In fact, these quantities will be conserved in any third quantized process, including the formation and evaporation 
of a black hole. Thus, this formalism can also be used to solve the information paradox in a black hole. 

In order to construct a third quantized Noether's theorem, we first some results in loop quantum cosmology \cite{1}-\cite{1a}.
In the loop quantum cosmology
 the curvature of $A^i_\mu$ is expressed through the holonomy around a loop and the  area of such a loop cannot 
cannot be smaller than a fixed minimum area because 
the smallest  eigenvalue of the area operator in loop quantum gravity is nonzero.  
Furthermore, the  eigenstates of the volume operator 
$\cal{V}$ are  
$ {\cal{V}} |\nu \rangle = 2 \pi \gamma G |\nu| |\nu \rangle
$, where $\nu = \pm a^2 {\cal{V}}_0 /2\pi \gamma G$ has the dimensions of length. 
So, in Planck units the Hamiltonian constrain for a homogeneous isotropic
 universe with a massless scalar field $\phi$,  can  be written as  \cite{a}-\cite{as}
\begin{equation} 
 K ^2\Phi(\nu, \phi) = [E^2 - \partial^2_\phi] \Phi(\nu, \phi) =0,
\end{equation}
where $\nu_0 =4$ and 
\begin{eqnarray}
 E^2 \Phi(\nu, \phi) &=& - [B(\nu)]^{-1}C^+(\nu) \Phi(\nu+4 , \phi)
 - [B(\nu)]^{-1}C^0(\nu)\Phi(\nu, \phi)  \nonumber \\ && - [B(\nu)]^{-1}C^-(\nu)\Phi(\nu-4, \phi).
\end{eqnarray}
Now, $K_1 = E$ and $K_2 = \partial_{\phi}$ and $\eta_{\mu\nu} =(1,-1)$, so we have $\eta^{\mu\nu}K_{\mu} K_{\nu}  = K^2$. 
Just as the wave function of first quantized theories is viewed as a classical field in second quantized formalism, the wave function of the 
second quantized theory is viewed as a classical field in third quantized formalism. Hence, the wave function of loop quantum cosmology 
will now be viewed as the classical field of group field cosmology. 
So, we can write the free part of the group field theory as 
\begin{equation}
 S_{b} = \sum_\nu \int d\phi \, \,    \Phi (\nu, \phi)  K^2  \Phi(\nu, \phi). 
\end{equation}
Thus, if for the third quantized action is invariant under some symmetry transformation $\delta$, 
then we will have a conserved current corresponding to this symmetry
\begin{eqnarray}
 J^{\mu} ( \nu, \phi) = 
\left[ \frac{ \partial \mathcal{L}_{b}  }{\partial K_{\mu} {\Phi(\nu, \phi)}    } 
\delta\, 
{\Phi(\nu, \phi)} \right], 
\end{eqnarray}
where 
\begin{equation}
S_{b}  =  \sum_{\nu} \int d\phi \,\,  \mathcal{L}_{b}. 
\end{equation}
Furthermore, we have, $
 K^{\mu}J_{\mu}(\nu, \phi) =0
$.
We can now used the fact that the matter field act like the time variable, and define a Noether's charge as 
\begin{eqnarray}
Q_{N} ( \phi) = \sum_\nu \,\, 
\left[ \frac{ \partial \mathcal{L}_{b}  }{\partial_\phi {\Phi(\nu, \phi)}   } 
\delta\, 
{\Phi(\nu, \phi)}\right].
\end{eqnarray}
Thus, even when the new universes will form in a multiverse, this third quantized Noether's charge will be conserved. 
Furthermore, this charge will also be conserved during any topological change, like the formation and evaporation of a  black hole. 
Thus, it seems that for quantum gravitational second quantized quantities are not conserved, but rather third quantized quantities are
conserved. 

Now, we can also construct a  a fermionic group field cosmology 
by using the analogy with  a regular  two dimensions theory fermionic \cite{susy}. Thus, 
  we define a fermionic field as
 $\Psi_a (\nu, \phi) = (\Psi_1(\nu, \phi), \Psi_2(\nu, \phi) )$.
The spinor indices are
 raised and lowered by the second-rank 
antisymmetric tensors $C^{ab}$ and $C_{ab}$, respectively, 
These second-rank 
antisymmetric tensors also satisfy $C_{ab}C^{cb} = \delta^c_a$.
So, we have $(\gamma^\mu)_{ab} = (\gamma^\mu)_a^c C_{cb} 
= (\gamma^\mu)_{ba}$ 
\cite{bsusy}. 
Now we  
define $K_\mu = (E, \partial_\phi), \eta^{\mu\nu} = (1, -1),  K_{ab} = (\gamma^\mu )_{ab}K_\mu$.
and  and $K^2 =  [E^2 - \partial^2_\phi]$.
Now we can write  the action for the fermionic group field cosmology as \cite{sgfc}
\begin{equation}
 S_{f} = \sum_\nu \int d\phi \, \,  \Psi^b(\nu, \phi) K_b^a \Psi_a(\nu, \phi).
\end{equation}
The classical equations of motion obtained from this action again generated the 
Hamiltonian constraint for the loop quantum gravity. 
We can now define a Noether's current for a fermionic field theory as 
\begin{eqnarray}
 J^{\mu} ( \nu, \phi) = 
\left[ \frac{ \partial \mathcal{L}_{f}  }{\partial K_{\mu} {\Psi_a(\nu, \phi)}    } 
\delta\, 
{\Psi_a(\nu, \phi)}\right], 
\end{eqnarray}
where 
\begin{equation}
S_{f}  =  \sum_{\nu} \int d\phi \,\,  \mathcal{L}_{f}. 
\end{equation}
We again have, $
 K^{\mu}J_{\mu}(\nu, \phi) =0
$.
So, we can again used the fact that the matter field act like the time variable, and define a Noether's charge as 
\begin{eqnarray}
Q_{N} ( \phi) = \sum_\nu \,\, 
\left[ \frac{ \partial \mathcal{L}_{f}  }{\partial_\phi {\Psi_a(\nu, \phi)}    } 
\delta\, 
{\Psi_a(\nu, \phi)} \right].
\end{eqnarray}
These charges will be conserved again even during the course of interaction of different universes in the multiverse. 

\section{Supersymmetry}
In this section we will analyse the supersymmetric group field cosmology. 
These  bosonic and the fermionic actions describe bosonic and the fermionic  universes in the multiverse. Hence, 
we can construct a supersymmetric gauge invariant multiverse. To do so we define, two complex scalar super-group fields
$\Omega(\nu, \phi, \theta) $ and $ \Omega^{\dagger}  (\nu, \phi, \theta)$ and a 
spinor super-group field $\Gamma_a (\nu, \phi, \theta)$. All these fields  are suitably contracted with generators of a Lie algebra,
$[T_A, T_B] = i f_{AB}^C T_C$, 
\begin{eqnarray}
   \Omega(\nu, \phi, \theta) &=&\Omega^A(\nu, \phi, \theta) T_A, \nonumber \\ 
\Omega^{\dagger}(\nu, \phi, \theta) &=& \Omega^{\dagger A}(\nu, \phi, \theta)  T_A \nonumber \\ 
\Gamma_a (\nu, \phi, \theta) &=& \Gamma_a^A (\nu, \phi, \theta)T_A. 
\end{eqnarray}
These superfields transform under infinitesimal gauge
 transformations as
\begin{eqnarray}
  \delta \Omega^A(\nu, \phi, \theta) &=&  if_{CB}^A\Lambda^C (\nu, \phi, \theta)\Omega^B(\nu, \phi, \theta) ,\nonumber\\
\delta \Omega^{A \dagger}(\nu, \phi, \theta)  &=& -i f^A_{CB}\Omega^{C\dagger}(\nu, \phi, \theta) \Lambda^B(\nu, \phi, \theta), \nonumber\\
 \delta \Gamma^A_a(\nu, \phi, \theta) &=& \nabla_a \Lambda^A(\nu, \phi, \theta).
\end{eqnarray}
 Now we define a super-covariant derivative of these superfields as
\begin{eqnarray}
  \nabla_a  \Omega^A(\nu, \phi, \theta)&=& D_a\Omega^A(\nu, \phi, \theta) -i f_{CB}^A\Gamma^C_a(\nu, \phi, \theta) \Omega^B(\nu, \phi, \theta),\nonumber\\
\nabla_a \Omega^{A \dagger}(\nu, \phi, \theta)  &=& D_a \Omega^{A \dagger}(\nu, \phi, \theta)  
+ i  f_{CB}^A\Omega^{C \dagger}(\nu, \phi, \theta)  \Gamma^B_a (\nu, \phi, \theta), 
\end{eqnarray}
where $ D_a = \partial_a + K^b_a \theta_b$.
We also define  $
 \omega_a (\nu, \phi, \theta) = \nabla^b \nabla_a \Gamma_b(\nu, \phi, \theta) $. 
We can write the action for the gauge theory as \cite{sgfc}
\begin{eqnarray}
S_{ga} &= &\sum_\nu \int d\phi \, \,    [D^2 [\Omega^{\dagger} ( \nu, \phi, \theta) \nabla^2 \Omega ( \nu, \phi, \theta) 
\nonumber \\ && \,\,\,\,\,\,\,\,\,\,\,\,\,\,\,\,\,\,\,\,\,\,\,\,\, + \omega^a ( \nu, \phi, \theta) \omega_a( \nu, \phi, \theta)  ]]_|. 
\end{eqnarray}
As this theory has a gauge symmetry, as it is invariant under the super-gauge transformations. 
We can now define a Noether's current for this supersymmetric field theory. We can then define a  Noether's charge for it as 
\begin{eqnarray}
Q_{N} ( \phi) &=& \sum_\nu \,\, 
\left[ \frac{ \partial \mathcal{L}_{ga}  }{\partial_\phi {\Omega ( \nu, \phi, \theta)}   } 
\delta\, 
{\Omega( \nu, \phi, \theta)}+
\frac{ \partial \mathcal{L}_{ga}  }{\partial_\phi {\Omega^{\dagger} ( \nu, \phi, \theta)}   } 
\delta\, 
{\Omega^{\dagger} ( \nu, \phi, \theta)} \right. \nonumber \\ &&\left.+
\frac{ \partial \mathcal{L}_{ga}  }{\partial_\phi {\Gamma_a(\nu, \phi, \theta)}   } 
\delta\, 
{\Gamma_a(\nu, \phi, \theta)}
\right],
\end{eqnarray}
where 
\begin{equation}
S_{ga}  =  \sum_{\nu} \int d\phi \,\,  \mathcal{L}_{ga}. 
\end{equation}

We need to fix a gauge before we can quantize it. Thus, we chose the following gauge \cite{sgfc}
\begin{equation}
 D^a \Gamma_a ( \nu, \phi, \theta) =0.
\end{equation}
This can be incorporated at a quantum level by adding the following gauge fixing term to original classical action, 
\begin{equation}
 S_{gf} = \sum_\nu \int d\phi \, \,    [D^2[ B ( \nu, \phi, \theta) D^a \Gamma_a ( \nu, \phi, \theta)]]_|, 
\end{equation}
where $B( \nu, \phi, \theta) = B^A( \nu, \phi, \theta) T_A$ is a 
auxiliary superfield. This field can be integrated out, in the path 
integral formalism, to impose the 
gauge fixing condition. Now we can write the ghost term by  gauge transforming the gauge fixing condition, replacing 
$\Lambda(\nu, \phi, \theta) = \Lambda^A (\nu, \phi, \theta) T_A$
 by ghost superfield $ {C}( \nu, \phi, \theta) =  {C}^A( \nu, \phi, \theta)  T_A$
and then  contracted it with anti-ghost superfield $ \overline{C}( \nu, \phi, \theta) = \overline{C}^A( \nu, \phi, \theta)  T_A$,   
\begin{equation}
  S_{gh} = \sum_\nu \int d\phi \, \,   
 [D^2[\overline C( \nu, \phi, \theta) D^a \nabla_a C 
( \nu, \phi, \theta)]]_|.
\end{equation}
Now the combined action 
$ S = S_{ga}+  S_{gh}+  S_{gf}$ is invariant under 
 the third quantized BRST transformations are given by
\begin{eqnarray}
  s\,\Omega^A(\nu, \phi, \theta)&= &if_{CB}
  ^AC^C (\nu, \phi, \theta)\Omega^B(\nu, \phi, \theta) , \nonumber \\
s\,  \Omega^{A \dagger}(\nu, \phi, \theta)  &= & -i f^A_{CB}\Omega^{C\dagger}(\nu, \phi, \theta) C^B(\nu, \phi, \theta), 
\nonumber \\ 
s\, C^A( \nu, \phi, \theta)&= &f^A_{CB}C^{C\dagger}(\nu, \phi, \theta) C^B(\nu, \phi, \theta), 
\nonumber \\
 s\, \Gamma^A_a ( \nu, \phi, \theta) &= & \nabla_a C^A( \nu, \phi, \theta), 
\nonumber \\ 
s\, \overline{C} ^A( \nu, \phi, \theta) &= & B^A( \nu, \phi, \theta),
 \nonumber \\
 s\, B^A( \nu, \phi, \theta) &= &0.
\end{eqnarray}
This action is also invariant under the following anti-BRST transformations. 
\begin{eqnarray}
 \overline s\,\Omega^A(\nu, \phi, \theta)&= &if_{CB}
  ^A \overline C^C (\nu, \phi, \theta)\Omega^B(\nu, \phi, \theta) , \nonumber \\
\overline s\,  \Omega^{A \dagger}(\nu, \phi, \theta)  &= & -i f^A_{CB}\Omega^{C\dagger}(\nu, \phi, \theta) \overline C^B(\nu, \phi, \theta), 
\nonumber \\ 
\overline s\, \overline C^A( \nu, \phi, \theta)&= &f^A_{CB}\overline C^{C}(\nu, \phi, \theta) \overline C^B(\nu, \phi, \theta), 
\nonumber \\
 \overline s\, \Gamma^A_a ( \nu, \phi, \theta) &= & \nabla_a \overline C^A( \nu, \phi, \theta), 
\nonumber \\ 
\overline s\, {C} ^A( \nu, \phi, \theta) &= & - B^A( \nu, \phi, \theta) + 
f^A_{CB}\overline C^{C}(\nu, \phi, \theta) C^B(\nu, \phi, \theta),
 \nonumber \\
 \overline s\, B^A( \nu, \phi, \theta) &= &f^A_{CB}\overline C^{C}(\nu, \phi, \theta) B^B(\nu, \phi, \theta).
\end{eqnarray}

 Both these transformations are nilpotent,
$s^2 =  \overline s^2 =0 $. In fact, they also satisfy, $ s\overline s + \overline s s = 0$.
The sum of the ghost term with the  gauge fixing term can  be 
written as a total third quantized BRST or a total third quantized anti-BRST variation
\begin{eqnarray}
  S_{gh} + S_{gf} &=& 
\sum_\nu \int d\phi \, \,    s \, [D^2 [\overline{C}  ( \nu, \phi, \theta) D^a \Gamma_a ( \nu, \phi, \theta) ]]_| \nonumber \\ 
&=& 
- \sum_\nu \int d\phi \, \,    \overline s \, [D^2 [{C}  ( \nu, \phi, \theta) D^a \Gamma_a ( \nu, \phi, \theta) ]]_|.
\end{eqnarray}
This can also be  be written as 
\begin{eqnarray}
   S_{gh} + S_{gf} &=& \frac{1}{2}
\sum_\nu \int d\phi \, \,   \overline s s \, [D^2 [\Gamma^a ( \nu, \phi, \theta)\Gamma_a  ( \nu, \phi, \theta) ]]_| \nonumber \\ 
&=& 
- \frac{1}{2}\sum_\nu \int d\phi \, \,    s \overline s \, [D^2 [\Gamma^a ( \nu, \phi, \theta) \Gamma_a ( \nu, \phi, \theta) ]]_|.
\end{eqnarray} 
 Thus, the sum of the gauge fixing term and the ghost term can be expressed as a combination of third quantized BRST and third quantized anti-BRST 
transformations.

\section{ Unitarity in the Multiverse}
Now in the multiverse the universe can collide with each other to form other universes. From 
the point of view of a single universe, this collision will appear as the big bang in which it was formed. 
It is possible that even our own universe formed because of the collision of two previous universes. 
In a second quantized theory, it is possible to show that when the BRST or the anti-BRST transformations are nilpotent, 
the resultant quantum field theory is unitarity. We have seen that the third quantized  BRST
 and the third quantized anti-BRST transformations are nilpotent. Thus,   
in this section we will show that creation and annihilation of universes in the multiverse is described by a 
unitarity process.  

We first note that it is possible  to construct  Norther's charges corresponding to the BRST and the anti-BRST symmetries 
of group field cosmology with a gauge symmetry. 
The Norther's charges corresponding to the BRST is given by 
\begin{eqnarray}
 Q_B  & = &   \sum_\nu \,\, 
\left[ \frac{ \partial \mathcal{L}_{eff}  }{\partial \partial_{\phi} {\Gamma_a } (\nu, \phi, \theta )   } 
 s\, 
{\Gamma_a }(\nu, \phi, \theta )   +
 \frac{ \partial \mathcal{L}_{eff}  }{\partial  \partial_{\phi} C (\nu, \phi, \theta )   }  s\, C (\nu, \phi, \theta )  
 \right.\nonumber \\ &&   +
\frac{ \partial \mathcal{L}_{eff}  }{\partial  \partial_{\phi} \overline{C} (\nu, \phi, \theta )   }
 s\, \overline{C}  (\nu, \phi, \theta )  
 +
\frac{ \partial \mathcal{L}_{eff}  }{\partial \partial_{\phi} B  (\nu, \phi, \theta )  }  s\, B (\nu, \phi, \theta ) 
\nonumber \\ && \left.  + 
\frac{ \partial \mathcal{L}_{eff}  }{\partial \partial_{\phi} \Omega (\nu, \phi, \theta )   }  s\, 
\Omega (\nu, \phi, \theta )    +
 \frac{ \partial \mathcal{L}_{eff}  }{\partial \partial_{\phi} \Omega^{ \dagger} (\nu, \phi, \theta )   } 
 s\, 
\overline \Omega^{ \dagger}  (\nu, \phi, \theta ) \right], 
\end{eqnarray}
and the Norther's charge corresponding to the anti-BRST is given by
\begin{eqnarray}
 Q_B  & = &   \sum_\nu \,\, 
\left[ \frac{ \partial \mathcal{L}_{eff}  }{\partial \partial_{\phi} {\Gamma_a }(\nu, \phi, \theta )   } 
 \overline{s} \, 
{\Gamma_a } (\nu, \phi, \theta )   +
 \frac{ \partial \mathcal{L}_{eff}  }{\partial  \partial_{\phi} C (\nu, \phi, \theta )   }  \overline{s} \, C (\nu, \phi, \theta )  
 \right.\nonumber \\ &&   +
\frac{ \partial \mathcal{L}_{eff}  }{\partial  \partial_{\phi} \overline{C} (\nu, \phi, \theta )   }
 \overline{s} \, \overline{C}  (\nu, \phi, \theta )  
 +
\frac{ \partial \mathcal{L}_{eff}  }{\partial \partial_{\phi} B  (\nu, \phi, \theta )  }  \overline{s} \, B (\nu, \phi, \theta ) 
\nonumber \\ && \left.  + 
\frac{ \partial \mathcal{L}_{eff}  }{\partial \partial_{\phi}  \Omega  (\nu, \phi, \theta )   }  \overline{s} \, 
  \Omega (\nu, \phi, \theta )    +
 \frac{ \partial \mathcal{L}_{eff}  }{\partial \partial_{\phi} \Omega^{ \dagger}  (\nu, \phi, \theta )   } 
 \overline{s} \, 
 \Omega^{ \dagger}  (\nu, \phi, \theta ) \right], 
\end{eqnarray}
where 
\begin{equation}
  S = \sum_{\nu} \int d\phi \,\, \mathcal{L}_{eff}. 
\end{equation}
It may be noted that we have again   used the fact that matter fields act like the time variable in this formalism.

 Now these  charge associated with
the BRST and the anti-BRST symmetries transformation commutes
 are  conserved. These charges 
 are also nilpotent
\begin{equation}
 Q_B^2 =  \overline{Q}_B^2 =0. 
\end{equation}
 Now we define physical states as states that are annihilated by $Q_B$,
$
 Q_B |\phi_p \rangle =0
$,
or as states that are annihilated by $\overline{Q}_B$,
$
 \overline{Q}_B |\phi_p \rangle =0
$.
The inner product of those
 physical states, which are obtained from unphysical states by the action of these charges, 
 vanishes with all other physical states.
Thus, all  physical information lies in the physical states which are not obtained by the action of these 
charges on unphysical states. 

 Now if the asymptotic physical states are $ |{out,a }\rangle$ and $ |{in, b}\rangle$,
 then a typical $\mathcal{S}$-matrix element can be written as
\begin{equation}
\langle{out}|{in}\rangle = \langle {pa}|\mathcal{S}^{\dagger}\mathcal{S}| {pb}\rangle.
\end{equation}
As  the states $\mathcal{S}|\phi_{pb}\rangle$ must be a linear combination of states physical states. 
 So we can write 
\begin{equation}
\langle{pa}|\mathcal{S}^{\dagger}\mathcal{S}|{pb}\rangle = \sum_{i}\langle {pa}|\mathcal{S}^{\dagger}|{0,i}\rangle
\langle{0,i}| \mathcal{S}|{pb}\rangle.
\end{equation}
Since the full $\mathcal{S}$-matrix is unitary this relation implies that the  $S$-matrix restricted to physical sub-space is also unitarity. 
Thus, the scattering of universes in a multiverse is a unitarity process. 

\section{Conclusion}
In this paper we first reviewed group field theory as third quantized loop quantum gravity. 
We analysed the properties of a   supersymmetric group field theory with a gauge symmetry. 
It was demonstrated that in  a multiverse different universes can collide and get annihilated.
Furthermore, new universes can form from the collision of these 
previous universe. So, a third quantized Noether's charge was constructed. This charge was conserved in the multiverse, 
even when the number of universes was not conserved. 
We also analysed a model of group field cosmology with gauge symmetry. 
This gauge symmetry was fixed to give rise to third quantized BRST and third quantized anti-BRST symmetries. 
The total Lagrangian  that was obtained by a sum of the original Lagrangian with the gauge fixing term 
and the ghost term was invariant under both these third quantized transformations. 
Furthermore, the third quantized Noether's charges, corresponding to the invariance of the group field 
cosmology under the third quantized BRST and third quantized anti-BRST transformations, was also constructed. 
These charges were used to demonstrated that the formation of universes in a multiverse is a unitarity process. 

It would be interesting to carry  this analyses for the 
super group field cosmology in  non-linear gauges. It is known that in this gauges the 
sum of the gauge fixing term and the ghost term can be expressed as a total BRST or a total anti-BRST variation 
in gauge theories \cite{gvsfad}. It will be interesting to analyse if a similar result holds here. 
It may be noted that  both the fermionic and bosonic group field theory have been studied and color has also been 
incorporated in group field theory \cite{color}-\cite{colora}. It would thus be interesting to
analyse a supersymmetric group field theory with gauge symmetry.
This way we will be symmetrizing the full group field theory. 
 We will be able to develop that theory in analogy with the usual gauge theory. 
It will be possible to derive this  theory in superspace formalism with manifest supersymmetry.

\section*{Appendix}
When ever the sum of the  
gauge fixing term and the ghost term can be written as 
a combination of the total BRST and the total anti-BRST variation, 
the total Lagrangian density is invariant under  a set of symmetry 
transformations which obey a
$SL(2, R)$ algebra called  Nakanishi-Ojima algebra \cite{gvsfad}. 
We will show that this algebra also hold for this group field cosmology. To do so we first 
define the following transformations,
\begin{eqnarray}
 \delta_{1}\, C^{A}(\nu, \phi, \theta) = 0,
&  & \delta_{1}\, B^A(\nu, \phi, \theta) =  f^A_{BC} C^{B}(\nu, \phi, \theta)  C^C(\nu, \phi, \theta), 
\nonumber \\
 \delta_{1}\,   \Gamma_a ( \nu, \phi, \theta) = 0, 
  & &\delta_{1}\, \overline C^{A}(\nu, \phi, \theta) =  C^{A}(\nu, \phi, \theta), 
\nonumber \\
 \delta_{1}\,\Omega^A(\nu, \phi, \theta) = 0, 
&&
  \delta_{1}\,  \Omega^{A \dagger}(\nu, \phi, \theta) =   0,
\nonumber \\ 
 \delta_{2}\, C^{A}(\nu, \phi, \theta) = 0, 
&&\delta_{2}\,  B^A(\nu, \phi, \theta)  =   f^A_{BC}\overline C^{B}(\nu, \phi, \theta) \overline C^C(\nu, \phi, \theta), 
\nonumber \\ 
 \delta_{2}\,   \Gamma_a ( \nu, \phi, \theta) = 0, 
 && \delta_{2}\, \overline C^{A}(\nu, \phi, \theta) =  C^{A}(\nu, \phi, \theta), 
\nonumber \\  \delta_{2}\,\Omega^A(\nu, \phi, \theta) = 0, 
&&  \delta_{2}\,  \Omega^{A \dagger}(\nu, \phi, \theta) =   0.
\end{eqnarray}
Now we can see that  these 
 transformations, the BRST transformation and the anti-BRST transformation 
 along with the $FP$-conjugation denoted by $\delta_{FP}$, form the 
Nakanishi-Ojima $SL(2, R)$ algebra,
\begin{eqnarray}
 [s,s]_{\star} =0, && [\overline{s},\overline{s}]_{\star} =0, \nonumber \\ 
{[s, \overline{s}]}_\star =0, && [\delta_{1}, \delta_{2}]_\star = - 2 \delta_{FP} \nonumber \\ 
{[\delta_{1}, \delta_{FP}]}_\star = -4 \delta_{1}, && [\delta_{2}, \delta_{FP}]_\star = 4 \delta_{2},\nonumber \\ 
{[s, \delta_{FP}]}_\star = - 2s , && [\overline s, \delta_{FP}]_\star = 2\overline s,\nonumber \\ 
{[s, \delta_{1}]}_\star = 0, && [\overline{s}, \delta_{1}]_\star = -2 s,\nonumber \\ 
{[s, \delta_{2}]}_\star = 2 \overline{s}, && [\overline{s}, \delta_{2}]_\star = 0.
\end{eqnarray}


\begin{thebibliography}{99}
\bibitem{1}A. Ashtekar,   Phys. Rev. D36, 1587 (1987)
\bibitem{svfdfkffkvfwpopop}
A. Ashtekar, Phys. Rev. Lett. 57,  2244 (1986)
\bibitem{pklkk}
C. Rovelli and L. Smolin, Nucl. Phys. B331, 80 (1990)
\bibitem{davfadlkk}
C. Rovelli, Quantum Gravity, Cambridge University Press, 
Cambridge, UK. (2007) 
\bibitem{10}G. Date and G. M. Hossain, 
SIGMA 8, 010 (2012) 
\bibitem{x}A. Ashtekar and A. Corichi,  Class. Quant. Grav. 20, 4473 (2003)
\bibitem{y}A. Ashtekar, Nature Physics 2, 726 (2006)
\bibitem{z}E. R. Livine and J. Tambornino, J. Math. Phys. 53, 012503 (2012)
\bibitem{1a}
M. Geiller, M. L. Rey and K.  Noui, 
Phys. Rev. D84, 044002 (2011) 
\bibitem{s}
C. Rovelli and L. Smolin, Phys. Rev. D53, 5743 (1995)
\bibitem{gcjgcgc}J. C. Baez, Adv. Math. 117, 253 (1996)
\bibitem{dwnsbdcsdbkcsdbkcdsdkscksd}
E. Bianchi, E. Magliaro and C. Perini, 
Phys. Rev. D82, 024012 (2010)
\bibitem{sa}C. Rovelli and  F. Vidotto, Phys. Rev. D81, 044038 (2010)
\bibitem{f}
L. Freidel and K. Krasnov,  Adv. Theor. Math. Phys. 2, 1183 (1999)
\bibitem{gcgcspinfoam}J. C. Baez,
Lect. Notes. Phys. 543, 25 (2000) 
\bibitem{ccdfcecefcfcdecdcspinfoam}A. Perez,
Class. Quant. Grav. 20, R43 (2003)
\bibitem{spinfomemmodelshvjh}
M. Geiller and K. Noui, Class. Quant. Grav. 29, 135008 (2012)
\bibitem{fa}V. Bonzom, Phys. Rev. D84, 024009 (2011)
\bibitem{t}A. Buonanno, M. Gasperini, M. Maggiore 
and  C. Ungarelli, Class. Quant. Grav. 14, L97 (1997)
\bibitem{dqfv}
L. O. Pimentel and C. Mora, Phys.Lett. A280 (2001) 191-196
\bibitem{fcbmvnmzxcvbds}
S. R. Perez and P. F. G. Diaz, Phys. Rev. D81, 083529 (2010) 127
\bibitem{ta}V. P. Maslov and O. Y. Shvedov, Phys. Rev. D60, 105012 (1999)
\bibitem{g}
A.  Baratin and D. Oriti, Phys. Rev. D 85, 044003 (2012)
\bibitem{hjhfjfhjk}
M. Smerlak, Class. Quant. Grav. 28, 178001 (2011)
\bibitem{kkfkkjkj}
A. Baratin, F. Girelli and D. Oriti, Phys. Rev. D83, 104051 (2011)
\bibitem{ga}A. Tanasa, J. Phys. A45,  165401 (2012)
\bibitem{c}
A. Ashtekar, J. Phys. Conf. Ser. 189, 012003 (2009)
\bibitem{sdndkdncomm}L. Qin, G. Deng and Y. Ma, Commun. Theor. Phys. 57, 326 (2012)
 \bibitem{sdfawe}B. Gupt and P. Singh,  Phys. Rev. D85, 044011 (2012)
\bibitem{ewqfaf}E. W. Ewing, Class. Quant. Grav. 29, 085005 (2012)
\bibitem{ca}L. Qin and Y. Ma, Mod. Phys. Lett. A27,  1250078 (2012)
\bibitem{m}G. Calcagni, S. Gielen and D. Oriti, 
Class. Quantum Grav. 29, 105005 (2012)
\bibitem{m2} L. A. Glinka, Grav. Cosmolo. 15, 317 (2009)
\bibitem{ma}S.  Gielen, J. Phys. Conf. Ser. 360,  012029 (2012)
\bibitem{sgfc}M. Faizal, Class. Quant. Grav. 29,  215009 (2012) 
\bibitem{brst}C. Becchi, A. Rouet and R. Stora, Phys. Lett. B52,  344 (1974)
\bibitem{acvds}C. Becchi, A. Rouet and R. Stora,  Ann. Phys. 98, 287 (1976) 
\bibitem{bkdashcabd}I.V. Tyutin, Lebe. Phys. Inst. preprint 39 (1975)
\bibitem{brst1}C. Becchi, A. Rouet and R. Stora, Commun. Math. Phys. 42,  127 (1975)
\bibitem{abrst}J. T. Mieg.  J. Math. Phys. 21, 2834 (1980)
\bibitem{csaddcsad} I. Y. Karataeva and S. L. Lyakhovich, Nucl. Phys. B545, 656 (1999) 
\bibitem{bakdhbc}M. Faizal, Found. Phys. 41, 270 (2011)
\bibitem{abrst1}P. Gregoire and M. Henneaux, Commun. Math. Phys. 157, 279 (1993) 
\bibitem{gauge}H. Min, T. Lee and P. Y. Pac,  Phys. Rev. D32, 440 (1985)
\bibitem{vdadsv}D. Dudal and H. Verschelde,  J. Phys. A36, 8507 (2003) 
\bibitem{gwfgvae}A. R. Fazio, V. E. R. Lemes, M. Picariello, M. S. Sarandy and S.P. Sorella, Mod. Phys. Lett. A18, 711 (2003)
\bibitem{gauge1}K. I. Kondo and T. Shinohara,  Phys. Lett. B491, 263
(2000)
\bibitem{a}
A. Ashtekar, T. Pawlowski, and P. Singh, Phys. Rev. D74, 084003 (2006) 
\bibitem{as}
A. Ashtekar, A. Corichi, and P. Singh, Phys. Rev. D77,
024046 (2008) 
\bibitem{color}
R. Gurau, Commun. Math. Phys. 304, 69 (2011) 
\bibitem{colora}
R. Gurau, Ann. Henri. Poincare. 11, 565 (2010) 
\bibitem{gvsfad}M. Faizal, Phys. Rev. D84, 106011 (2011)
\bibitem{susy}S. J. Gates Jr, M. T. Grisaru, M. Rocek and W. Siegel, Front.Phys. 58,  1 (1983) 
\bibitem{bsusy} D. V. Belyaev and P. V. Nieuwenhuizen, JHEP. 0804, 008 (2008)

\end{thebibliography}
\end{document}